\documentclass[twocolumn,twocolappendix]{aastex701}

\usepackage{graphicx}

\begin{document}

\title{A Major Geomagnetic Storm in 2024 October Linked to Sympathetic CME--Prominence Eruptions}

\author[orcid=0000-0001-5205-1713, sname='Wang']{Rui Wang}
\affiliation{State Key Laboratory of Solar Activity and Space Weather, National Space Science Center, Chinese Academy of Sciences, Beijing 100190, People's Republic of China}
\affiliation{School of Astronomy and Space Sciences, University of Chinese Academy of Sciences, Beijing 100049, People's Republic of China}
\affiliation{Key Laboratory of Space Weather, National Satellite Meteorological Center (National Center for Space Weather), China Meteorological Administration, Beijing 100081, People's Republic of China}

\email[show]{rwang@swl.ac.cn}  

\author[orcid=0000-0001-8188-9013, sname='Hu']{Huidong Hu}
\affiliation{State Key Laboratory of Solar Activity and Space Weather, National Space Science Center, Chinese Academy of Sciences, Beijing 100190, People's Republic of China}
\email[show]{huhd@nssc.ac.cn} 

\author[orcid=0000-0002-4016-5710, sname=Zhao]{Xiaowei Zhao}
\affiliation{Key Laboratory of Space Weather, National Satellite Meteorological Center (National Center for Space Weather), China Meteorological Administration, Beijing 100081, People's Republic of China}
\email[show]{zhaoxw@cma.gov.cn}

\author[orcid=0000-0002-2316-0870]{Chong Chen}
\affiliation{School of Microelectronics and Physics, Hunan University of Technology and Business, Changsha 410205, People's Republic of China}
\email{chenc@hutb.edu.cn}

\author[orcid=0000-0002-5431-6065]{Suli Ma}
\affiliation{State Key Laboratory of Solar Activity and Space Weather, National Space Science Center, Chinese Academy of Sciences, Beijing 100190, People's Republic of China}
\affiliation{School of Astronomy and Space Sciences, University of Chinese Academy of Sciences, Beijing 100049, People's Republic of China}
\email{masuli@nssc.ac.cn}

\author[orcid=0000-0002-1509-1529, sname='Yang']{Zhongwei Yang}
\affiliation{State Key Laboratory of Solar Activity and Space Weather, National Space Science Center, Chinese Academy of Sciences, Beijing 100190, People's Republic of China}
\affiliation{School of Astronomy and Space Sciences, University of Chinese Academy of Sciences, Beijing 100049, People's Republic of China}
\email{zwyang@swl.ac.cn}

\author[orcid=0000-0002-3032-6066]{Lei Lu}
\affiliation{Key Laboratory of Dark Matter and Space Astronomy, Purple Mountain Observatory, Chinese Academy of Sciences,
Nanjing 210033, People's Republic of China}
\email{leilu@pmo.ac.cn}

\author[orcid=0000-0003-4655-6939]{Li Feng}
\affiliation{Key Laboratory of Dark Matter and Space Astronomy, Purple Mountain Observatory, Chinese Academy of Sciences,
Nanjing 210033, People's Republic of China}
\email{lfeng@pmo.ac.cn}

\author[orcid=0009-0005-3941-1514]{Wenshuai Cheng}
\affiliation{State Key Laboratory of Solar Activity and Space Weather, National Space Science Center, Chinese Academy of Sciences, Beijing 100190, People's Republic of China}
\affiliation{School of Astronomy and Space Sciences, University of Chinese Academy of Sciences, Beijing 100049, People's Republic of China}
\email{chengwenshuai22@mails.ucas.ac.cn}

\author[orcid=0009-0004-3693-3141]{Chong Huang}
\affiliation{School of Electronic Information and Electrical Engineering, Huizhou University, Huizhou, 516001, People's Republic of China}
\affiliation{Guangdong Provincial Key Laboratory of Electronic Functional Materials and Device, Huizhou, 516001, People's Republic of China}
\email{huangchong@hzu.edu.cn}

\author[orcid=0000-0003-3142-217X]{Quan Wang}
\affiliation{National Astronomical Observatories, Chinese Academy of Sciences, Beijing 100101, People's Republic of China}
\affiliation{School of Astronomy and Space Sciences, University of Chinese Academy of Sciences, Beijing 100049, People's Republic of China}
\email{wangquan@nao.cas.cn}

\author[sname='Zhu']{Xiaoshuai Zhu}
\affiliation{State Key Laboratory of Solar Activity and Space Weather, National Space Science Center, Chinese Academy of Sciences, Beijing 100190, People's Republic of China}
\affiliation{School of Astronomy and Space Sciences, University of Chinese Academy of Sciences, Beijing 100049, People's Republic of China}
\email{zhuxiaoshuai@nssc.ac.cn}

\author[orcid=0000-0001-6306-3365]{Bei Zhu}
\affiliation{Space Engineering University, Beijing 101416, People's Republic of China}
\email{zhub@hgd.edu.cn}

\author[orcid=0009-0004-4832-0895]{Yiming Jiao}
\affiliation{State Key Laboratory of Solar Activity and Space Weather, National Space Science Center, Chinese Academy of Sciences, Beijing 100190, People's Republic of China}
\affiliation{School of Astronomy and Space Sciences, University of Chinese Academy of Sciences, Beijing 100049, People's Republic of China}
\email{jiaoyiming22@mails.ucas.edu.cn}


\begin{abstract}
Improving predictions of the geomagnetic impact of coronal mass ejections (CMEs) requires understanding how solar source properties relate to in-situ measurements at Earth. However, major geomagnetic storms frequently arise from interacting CMEs, complicating the link back to their solar origins. We analyze a CME interaction event that caused a major geomagnetic storm in 2024 October 10-11 (D$_{st}$ $\sim$-333 nT). Multiviewpoint observations reveal that the storm was related to a sympathetic eruption involving a quiescent filament and an active-region CME. The coronagraph on board the Advanced Space-based Solar Observatory clearly shows that this sympathetic eruption resulted in two distinct CMEs. Due to the overlap of the CMEs in the coronagraph field of view (FOV), a spheroid shock model was used to fit the observed shock. Kinematic analysis indicates that the interacting CMEs had completed their impulsive acceleration phase before entering the coronagraph FOV, with a slow deceleration continuing beyond 100 R$_\odot$. In-situ measurements indicate that the enhanced southward magnetic fields, arising from compression during CME interactions, were the primary driver of the storm. Compared to photospheric fields, the in-situ magnetic fields suggest that the trailing CME maintained flux-rope-like signatures consistent with the source region. In contrast, the compressed leading CME displayed varying magnetic configurations between Wind and STEREO-A, featuring distorted flux-rope signatures and inconsistent inferred axis orientations. Our study bridges solar source dynamics to in-situ multipoint measurements, providing key insights for space weather prediction. Nevertheless, the direct linkage between source-region magnetic field configurations and these measurements remains tentative and requires further investigation.

\end{abstract}



\section{Introduction} 

A key challenge in coronal mass ejections (CMEs) or interplanetary CMEs (ICMEs) forecasting is the prediction of the southward magnetic field (B$_z$) prior to arrival, as the geoeffective dawn--dusk electric field (E = -v$\times$B$_z$) critically depends on both the solar wind speed and the interplanetary magnetic field orientation. Previous studies indicate that large geomagnetic storms can be predicted from early signs in eruption source regions \citep[e.g.,][]{2013Liuxying,2015Liuxying,2016syntelis,2016Rui2}. However, CMEs are often not isolated eruptions but occur as multiple, simultaneous, or sequential eruptions, leading to complex interactions in interplanetary space \citep[e.g.,][]{2017lugaz,2017shenf,2024Chiyt}. These interactions result in the formation of highly intricate ejecta structures \citep{2024liuxying, 2024Rui2}. On the other hand, CME rotation \citep{2018LiuAyi, 2019Chenc} and deflection \citep{2004Cremades,2009Gopalswamy,2011Liewer} further complicate geoeffective predictions.

Previous studies have revealed that CMEs often exhibit a sequential nature, particularly in the form of homologous CMEs \citep[e.g.,][]{1984Woodgate, 2002zhangjun, 2022rui, 2024liuxying,2024Rui2}. These homologous CMEs may originate from the same polarity inversion line (PIL) within a single active region (AR) or, more commonly, from different PILs within the same AR, significantly increasing the likelihood of CME interactions. Furthermore, nearby solar eruptions can trigger additional AR eruptions or filaments, leading to sympathetic CMEs \citep[e.g.,][]{2011torok,2016Rui2}. These sympathetic CMEs are likely to cause successive eruptions with close angular and temporal separations, further enhancing the frequency and complexity of CME interactions.

Interactions between CMEs/ICMEs are highly complex, as these interactions often lead to the formation of complex ejecta \citep{2002Burlaga} or compound streams \citep{1987Burlaga}. \citet{2001gopalswamy} termed the formation of complex ejecta as ``CME-cannibalism.'' This involves not only ejecta--ejecta interactions but also interactions of ejecta with shocks. A CME-driven shock propagating into a preceding ICME can result in either a shock-in-ICME \citep{1987Burlaga,1997Vandas,2014Liuxying2} or an ICME-in-sheath structure \citep{2020Liuxying}. Both configurations enhance the magnetic field, speed, and density, potentially causing intense geomagnetic storms. Additionally, \citet{2014Liuxying1} proposed the ``perfect storm'' scenario, which emphasized that the leading CME ``preconditions'' the solar wind environment for subsequent CMEs. It has been proposed that the preconditioning effect of an earlier CME on the upstream solar wind constitutes a form of indirect interaction with a later CME \citep{2019Liuxying}. In summary, CME interactions significantly increase the complexity of CMEs. As concluded by \citet{2002Burlaga}, the nonlinear and irreversible merging of CMEs results in a loss of memory of their source conditions. This, in turn, makes the prediction of their geoeffectiveness considerably more challenging.

Many studies on ICME interactions neglect the conditions at their solar sources, whereas others focus on solar observations for space weather forecasting without accounting for the possibility of CME interactions. Linking geomagnetic effects from CME interactions to solar origins is challenging, as interactions of multiple CMEs complicate the identification of their source signatures \citep[e.g.,][]{2001gopalswamy,2012Lugaz,2014Liuxying1,2017lugaz}. This letter examines a CME interaction event, connecting CME sources from multiview with multipoint in-situ observations to uncover their intrinsic connections.

\section{Observations and Results}\label{s1}

At the onset of 2024 October, solar activity reached exceptionally high levels near the peak of Solar Cycle 25. Coronagraph observations from Solar Heliospheric Observatory/LASCO and STEREO-A/COR2 show what appears to be a single major halo CME erupting on October 9 and propagating toward Earth. We rule out the possibility of interactions with the CMEs from the previous days that were temporally and spatially unrelated (for more details, see Appendix A). On Oct-9, the only significant eruption was an X1.8 flare produced by AR 13848 at approximately 01:56 UT. Prior to the flare, a large-scale quiescent filament (hereafter QF) located west of the AR was connected to its vicinity (Figure 1a). The ascent of the filament preceded the eruptions in the AR, or rather, it triggered the AR eruptions, namely a sympathetic eruption event. See the animation of the rising QF in multiple extreme-ultraviolet (EUV) wavelengths. The red dashed lines in the top panel of Figure 1b (EUVI 195 \AA) indicate the adjacent legs of two filaments, which were anchored in opposite magnetic polarities. The western filament exhibited initial instability and approached the eastern filament, leading to magnetic reconnection and the formation of a larger-scale filament, as shown by the green dashed lines in Figure 1a. The western filament lies just below the torus-unstable region \citep[see the magenta contours in Figure 1c;][]{2006Kliem}, with a decay index of 1.5 at a critical height of $\sim$50 Mm, where the magnetic field was derived via the potential field source surface (PFSS) model. The Chinese H$\alpha$ Solar Explorer \citep[CHASE;][]{2022LiC}/H$\alpha$ imaging suggests an east-west filament axis with left-handed magnetic helicity, inferred from the right-bearing barb and surrounding magnetic field distributions \citep{1998Martin}.

\begin{figure*}[ht!]
\includegraphics[scale=0.8]{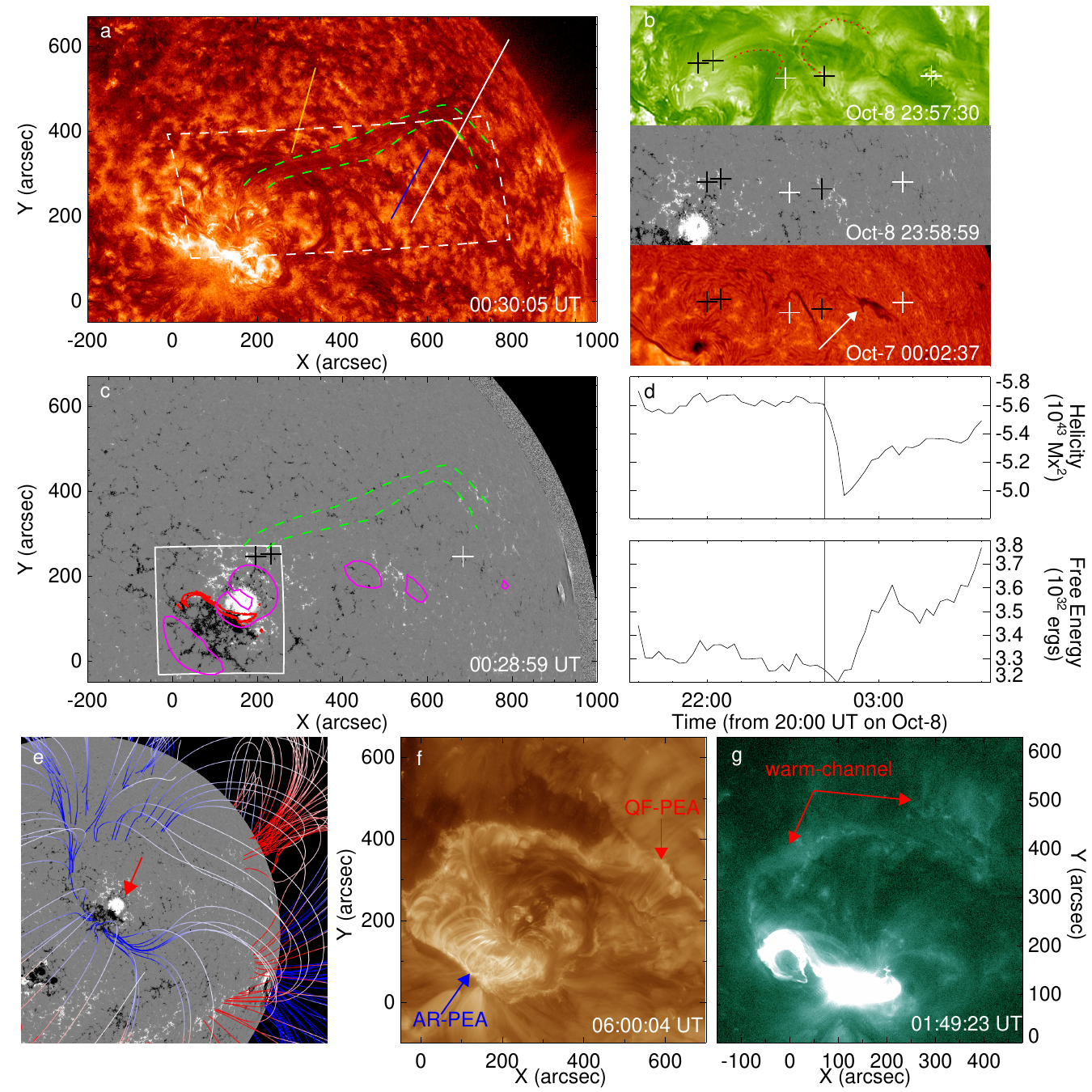}
\caption{Multiwavelength solar observations of the sympathetic eruption on 2024 Oct-9. (a) Rising QF (green) at AIA 304 \AA~. Colored slits indicate positions used for time--distance analysis of the QF rising motion and flare ribbon expansions. (b) EUVI 195 \AA~image showing the region of filament formation (dashed quadrilateral in panel (a)). Footpoints of the two reconnected filament segments are marked by black and white crosses on the coaligned HMI magnetogram and CHASE H$\alpha$ image; the white arrow points to the filament barb. (c) HMI magnetogram of AR 13848 and the nearby region. The white square outlines the domain used for magnetic field extrapolation in panel (d); red contours mark the brightened sigmoid structure at AIA 94 \AA; magenta contours show the decay index n = 1.5 isosurface at a height of $\sim$50 Mm. (d) Time evolution of magnetic helicity and free energy of the AR. (e) Large-scale magnetic fields by PFSS surrounding the AR. (f) PEAs associated with the two CMEs at AIA 193 \AA. (g) Inverted $\delta$-shaped warm-channel structure at AIA 94 \AA. An animation of panel (a) shows the QF rising process in multiwavelength AIA observations (01:38-02:53 UT, 19 s duration).}
\end{figure*}

Figure 1c shows a nonpotential sigmoid structure observed by the Atmospheric Imaging Assembly \citep[AIA;][]{2012Lemen} on board the Solar Dynamics Observatory \citep[SDO;][]{2012Pesnell} in the 94 \AA~(red contours), lying almost horizontally from west to east across the AR. We investigated the nonpotentiality of this AR. Using Space-weather Helioseismic and Magnetic Imager \citep[HMI;][]{2012Scherrer,2012Schou} Active Region Patches \citep[SHARPs;][]{2014Bobra,2014Hoeksema} as the boundary condition (white square in Figure 1c), we applied an optimization method \citep{2004Wiegelmann} to extrapolate the coronal magnetic field for a 10-hour sequence. Figure 1d shows that the magnetic energy did not exhibit a significant decrease during the eruption. The magnetic helicity calculation \citep{1984Berger,2013Yangsb,2018Yangsb} revealed a rapid decline during the eruption. According to the free energy--helicity diagram by \citet{2012tziotziou}, the energy and helicity magnitudes of this AR are characteristic of an eruptive AR, and it indeed produced an X1.8 flare. However, based on the statistics of \citet{2023wangq}, the energy ($\sim$1.8\%) and helicity change rate ($\sim$11\%) in Figure 1d indicate that this AR falls below the median for eruptive ARs, exhibiting relatively small change rates, with magnetic energy release far below the average level. Figure 1e shows the AR located outside the closed fields of a streamer belt and a coronal hole north of the AR. After the eruption, coronal dimmings appeared near the coronal hole (Figure 1f). AIA observations at 193 \AA~also show two posteruptive arcades (PEAs) in the AR and the filament source region. Figure 1g and the animation of Figure 1a show a large-scale warm-channel structure \citep{2022songhq} rising along an inverted $\delta$-shaped PIL, but no individual ejecta were observed departing from the sigmoid structure region.

\begin{figure*}[ht!]
\includegraphics[scale=0.7]{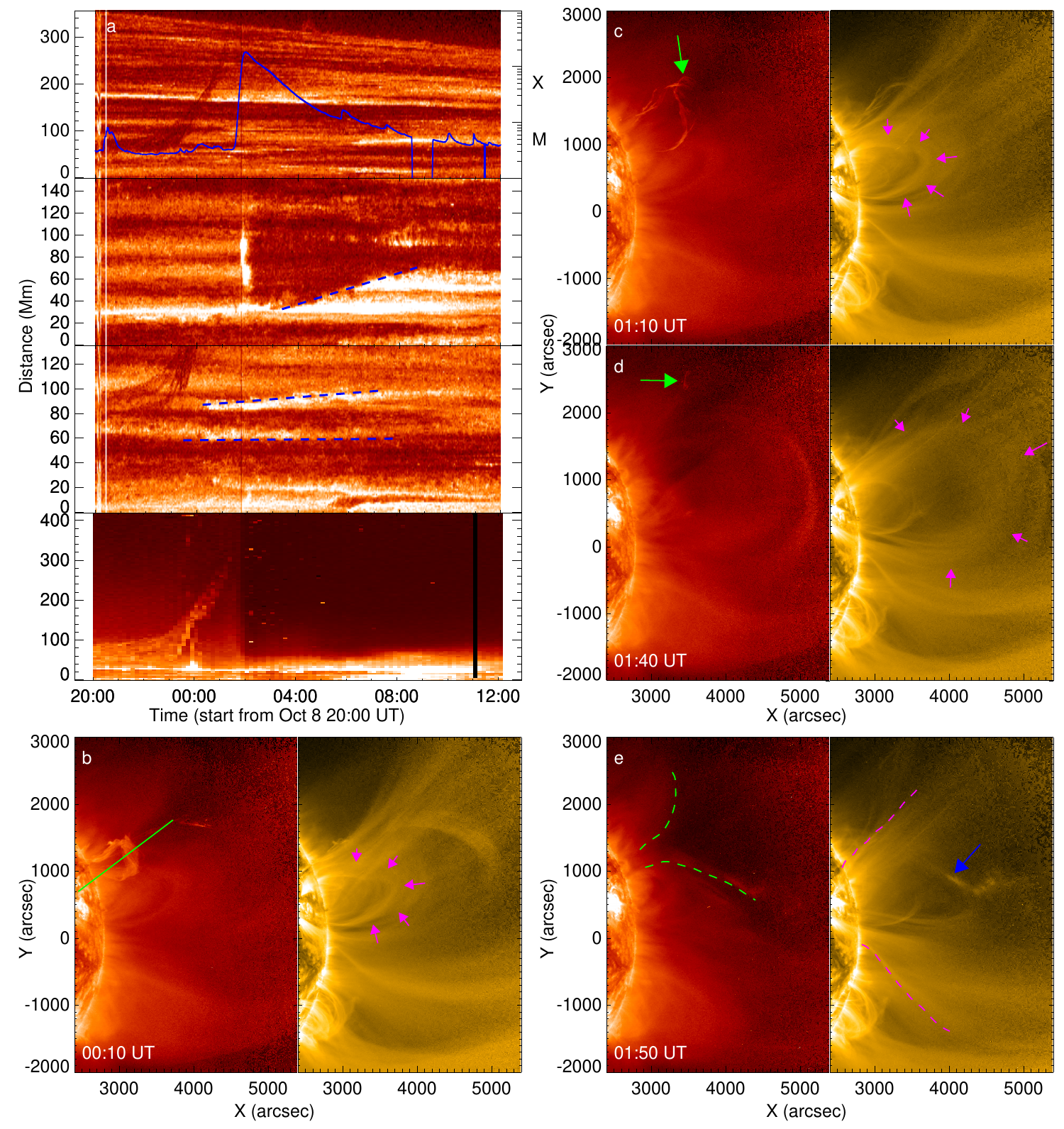}
\caption{Off-limb observations of the rising filament (prominence) by SolO. (a) Time--distance plots from the slits of Figures 1(a) and (b). GOES X-ray flux is overplotted. The blue dashed lines indicate the separation motions of flare ribbons. (b)-(e) Evolution of the filament (indicated by green arrows) by EUI/FSI 304 \AA~and 174 \AA. The magenta arrows indicate the expansion of the overlying loop above the AR. The blue arrow marks the coronal loops distorted by the filament eruption. An animation of panels (b)-(e) that showcases the QF and AR eruptions in the FSI from 00:00 to 03:00 UT (1 s duration).}
\end{figure*}

We placed three slits on the 304 \AA~image in Figure 1a: a white slit along the direction of the rise of the QF (the time--distance plot shown in the first panel in Figure 2a), a yellow slit along the direction of the expansion of the flare ribbons associated with the northern part of the inverted $\delta$-shaped PIL (second panel), and a blue slit along the direction of separation motions of the ribbons beneath the QF (third panel). The rise of the QF clearly preceded the AR eruptions. When the flare ribbons began to appear around 01:56 UT, the QF had already risen to a height of $\sim$380 Mm at 01:30 UT \citep[0.546 R$_\odot$; determined through stereoscopic measurements; see][]{2024Rui1} and had subsequently disappeared from the upper-right corner of the on-disk field of view (FOV) by 01:50 UT. Obviously, the flare ribbons were formed due to the eruption of the AR CME (hereafter AR-CME), stretching the overlying coronal fields (as shown by the PFSS fields in Figure 1e), independent of the QF, whereas the double ribbons associated with the QF just formed as the QF rose to a lower height. The AR-PEAs and QF-PEAs in Figure 1f were still observable until 17:00 UT on Oct-9, with faint signatures persisting into Oct-10. Their durations are consistent with the average EUV emission lifetime of PEAs reported by \citet{2004Tripathi}

Notably, this event was also observed by the Solar Orbiter \citep[SolO;][]{2020Muller}, as shown in Figure 6b. Positioned on the eastern side of the Sun with a 70$^\circ$ separation from Earth, SolO provided a side-on perspective of both the AR eruptions and the QF. Figures 2b-2e show a sequence of images captured by SolO's EUV Imager \citep[EUI;][]{2020Rochus} Full Sun Imager (FSI) during the eruptions, complemented by an animation showing the full dynamic evolution. Figure 2b captures the moment when the QF, formed after the reconnection in Figure 1b, begins to rise. The time-distance plot along the slit direction in Figure 2b has been converted to the on-disk filament timing. Figure 2c reveals a distinct kink structure in the QF. In Figure 2d, the coronal loops at 174 \AA~begin to expand outward rapidly, while the top structure of the QF remains faintly visible at 304 \AA, indicating that the filament has not yet been ejected. Figure 2e shows both structures have erupted. In the FSI 304 \AA~image, a cusp structure is visible after the eruptions (green dashed curves), just below the QF. In contrast, the center of the opened magnetic loops at 174 \AA~aligns with the AR eruptions (magenta dashed curves), which suggests that the QF and the AR eruptions occurred separately. Due to the low temporal resolution of FSI (10 minutes), further detailed analysis of this process is limited. However, considering the initial altitudes of the eruption, the CME associated with the QF (hereafter QF-CME) likely preceded the AR-CME, with a time difference of less than 10 minutes (constrained by the temporal resolution of FSI). Additionally, we observed that the lateral expansion of the QF-CME exerted a compressive effect on nearby coronal loops, which were not in the same plane of sky as the opened loops, further supporting the idea that two distinct CMEs were produced.

\begin{figure*}[ht!]
\includegraphics[scale=0.9]{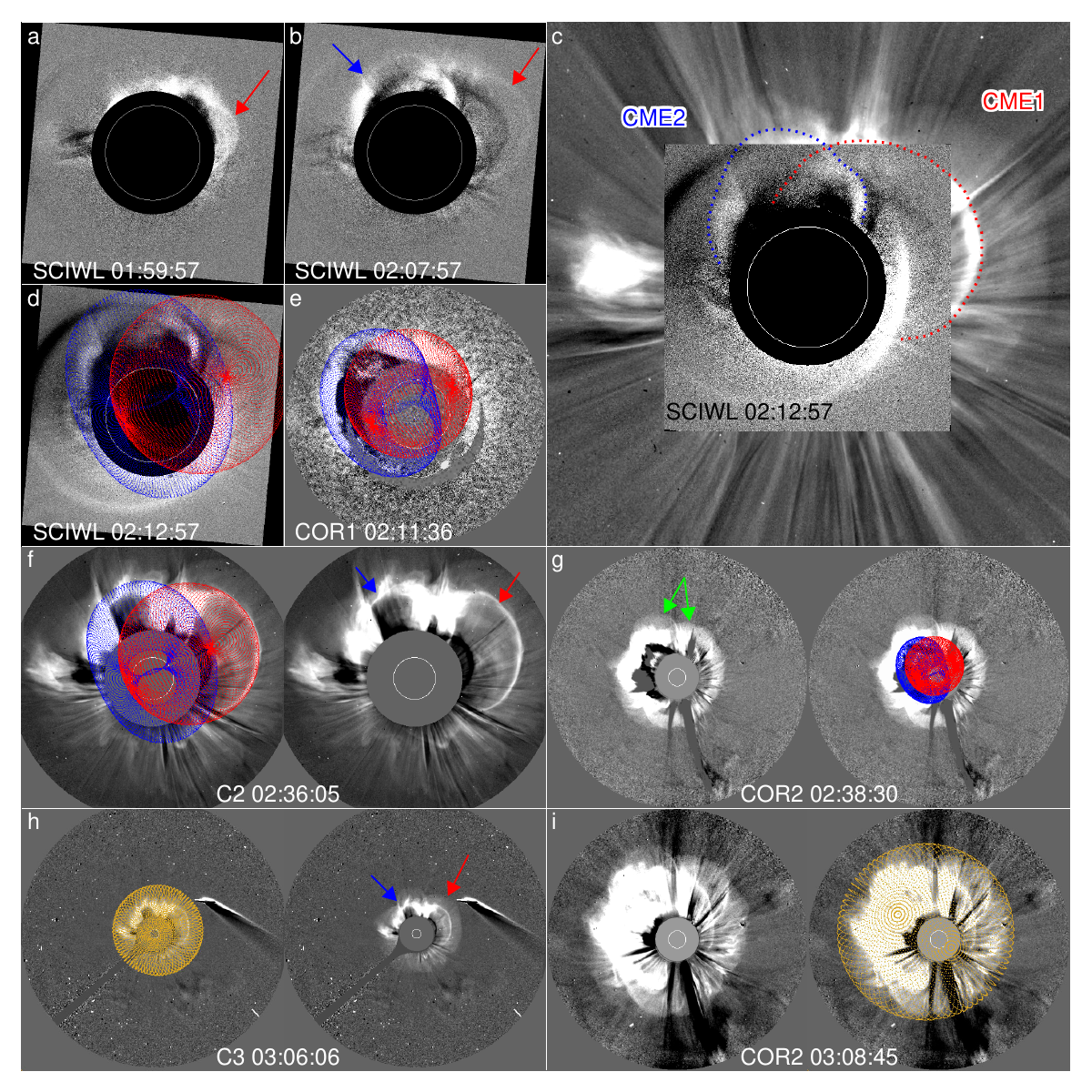}
\caption{Evolution of two CMEs observed by coronagraphs on Oct-9. (a) and (b) Running-difference (RD) white-light images observed by ASO-S/SCIWL. Red and blue arrows indicate CME1 (QF-CME) and CME2 (AR-CME), respectively. (c)
Composite RD images of SCIWL and LASCO/C2. Red and blue dotted lines outline the leading edges of two CMEs. (d) and (e) GCS wireframes overplotted on SCIWL and STEREO-A/COR1. Blue (red) wireframe corresponds to AR-CME (QF-CME). (f) and (g) GCS modeling from C2 and STEREO-A/COR2. The green arrows indicate two distinct shocks. (h) and (i) Spheroid shock modeling (golden) from LASCO/C3 and COR2. An animation of panels (a) and (b) shows two CMEs with distinct leading edges propagating outwards in the SCIWL from 01:57 to 02:55 UT (2 s duration).}
\end{figure*}

Figure 3 illustrates the evolution of the CMEs in coronagraph imagery following the eruptions. Figures 3a and 3b present observations from the Solar Corona Imager in White Light \citep[SCIWL;][]{2019FengLi,2019LiHui} on board the Advanced Space-based Solar Observatory \citep[ASO-S;][]{2022GanWQ}. SCIWL images the solar corona from 1.1 to 2.5 solar radii, offering critical observations on early CME dynamics and effectively bridging the observational gap between other coronagraphs and EUV imagers. In Figure 3a, a structure first emerges from the western limb at 01:59 UT, followed by another structure appearing northeast of the polar region (refer to the animation of Figure 3a). Spatially, the western ejecta correspond to the QF-CME, while the eastern CME should be the AR-CME. The QF-CME indeed preceded the AR-CME, which is consistent with the initial altitudes of the filament eruption stated above. The composite image in Figure 3c reveals two distinct leading edges (dotted lines) around 02:12 UT, indicating these are separate CMEs. A clear shock is already visible in the west. The Graduated Cylindrical Shell \citep[GCS;][]{2006Thernisien} wireframes in Figure 3d align well with the two CMEs outlined by the dotted lines in Figure 3c. These results indicate that the two CMEs overlap significantly, with the AR-CME dominating most of the ejecta in the COR1 FOV, while the QF-CME appears only as a portion on the right side (Figure 3e). Comparing views from C2 and COR2, although the two CMEs occupy the two ``wings'' of a butterfly-shaped structure in C2 (Figure 3f), the QF-CME is visible only as a small section on the right in COR2 (Figure 3g), where the left side is overwhelmingly dominated by the AR-CME. In COR2, the left side fits less accurately than the right because the GCS model is timed to later C2, and the left AR-CME expands faster Earthward than the right QF-CME along the line of sight. The green arrows in Figure 3g indicate two distinct shocks generated by the two CMEs, and these shocks overlap precisely where the two GCS structures intersect.

However, due to possible deformation caused by CME interactions, this fitting approach may introduce considerable theoretical uncertainty. In contrast, their associated shocks can be modeled using a simple spheroid model. For this purpose, we also employed a technique integrated into the ray-tracing code \citep{2006Thernisien}, which successfully reproduces the shock structure from both observational viewpoints. The model yields a propagation direction of W10$^\circ$, N16$^\circ$ and an initial linear speed of $\sim$1730 km s$^{-1}$. The two CMEs constituted a twin-CME scenario \citep{2012ligang}. The strongest solar proton event of Solar Cycle 25 was detected on Oct-9, with a peak flux of 1150 pfu at $>$10 MeV \footnote{\url{http://www.sepc.ac.cn/}}, likely driven by early coronal interactions between the two CMEs.

\begin{figure*}[ht!]
\plotone{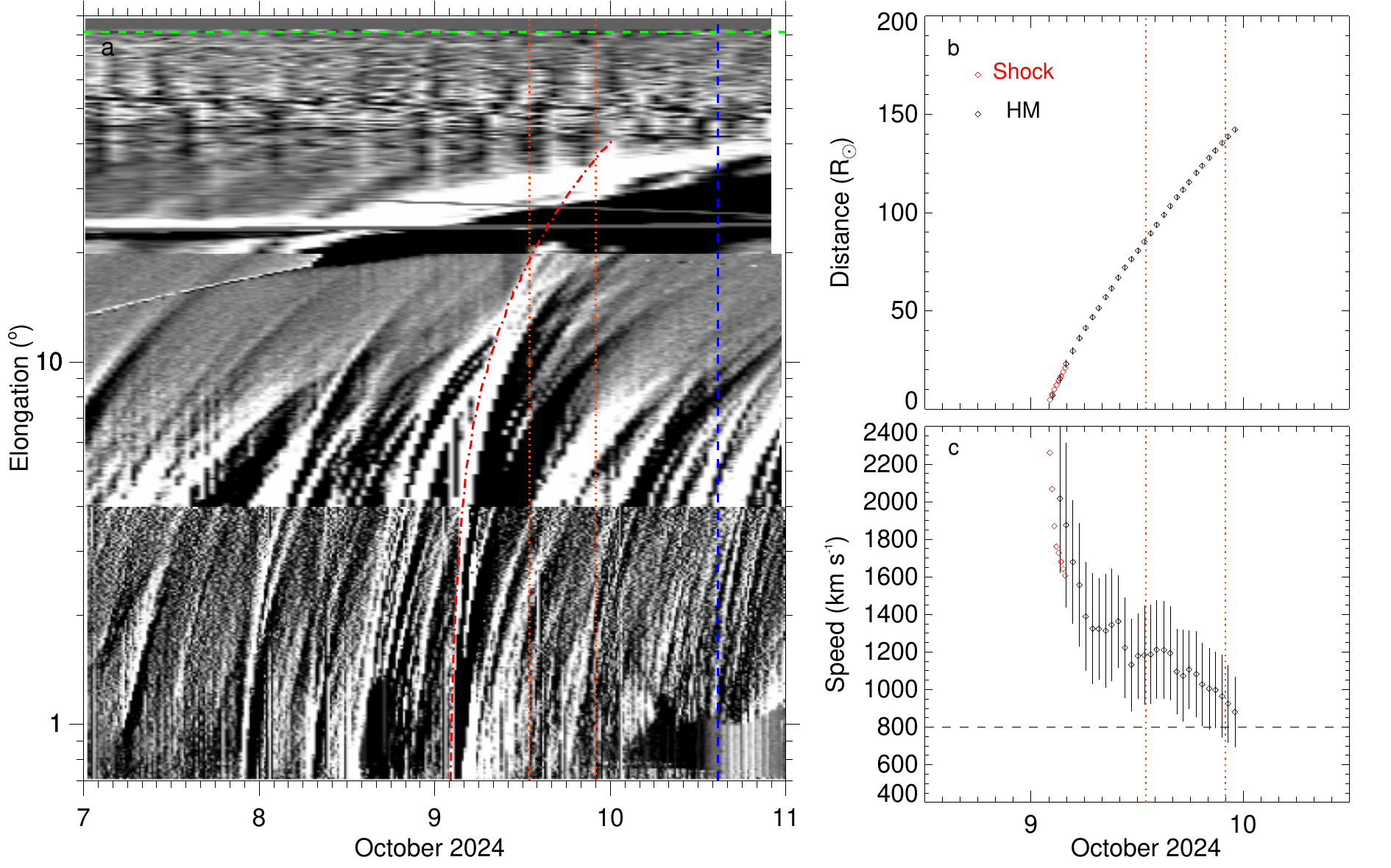}
\caption{Kinematics of the Oct-9 CME. (a) Time-elongation map constructed from RD images of COR2, HI1, and HI2 from STEREO-A along the ecliptic plane. The red curves indicate the CME track. The green horizontal dashed line denotes the elongation angle of the Earth. The blue vertical line shows the arrival time of the CME shock at the Earth. (b) Radial distance and (c) speed profiles of the CME leading edge derived from the spheroid shock model (red) and the HM approximation (black). The horizontal dashed line in (c) represents the shock speed of about 800 km s$^{-1}$ at Wind. Orange dotted lines bracket the comet-obscured area.}
\end{figure*}

We determined the CME kinematics from the results of the shock modeling as well as from the harmonic mean (HM) approximation \citep{2009Lugaz}, which assumes a spherical front attached to the Sun and moving along a fixed radial direction. For this, we used the elongation angles from the time-elongation map (see Figure 4a) and the averaged longitude (W10$^\circ$) as input to the HM approximation to derive the kinematic properties of the CME (Figure 4b and 4c). It indicates that the CMEs had already completed their impulsive acceleration phase before entering the coronagraph FOV. The peak speed of the shock nose is $\sim$2260 km s$^{-1}$. After entering the coronagraph, the CME underwent rapid deceleration inside $\sim$20 R$_\odot$, with speed dropping to $\sim$1200 km s$^{-1}$ by $\sim$100 R$_\odot$. This behavior is consistent with the dominance of aerodynamic drag during propagation for fast CMEs (initial speeds $>$ 900 km s$^{-1}$), as reported in the studies of \citet{2015sachdeva,2017sachdeva}, who showed that drag effects can become significant as early as 3.5-4 R$_\odot$. Comet interference prevented accurate measurement of the CME speed in HI2 (between the orange dotted lines). However, later analysis of in-situ data at Wind reveals a shock speed of $\sim$800 km s$^{-1}$. The trends in Figures 4a and 4c suggest that the CME is expected to stabilize near this value by 1 au, indicating that it was still far from its final speed by $\sim$100 R$_\odot$, with deceleration likely ongoing until $\sim$100-150 R$_\odot$. While \citet{2019Zhaoxw} found that deceleration ceased between $\sim$60 and 130 R$_\odot$ for the events they studied, prolonged deceleration beyond 100 R$_\odot$ occurred only in events involving CME interactions. \citet{2017Liuxying} also proposed that merged CMEs experience reduced deceleration due to increased mass and inertia.

Figure 5a shows the in-situ measurements at Wind from Oct-10 to 12. A forward shock arrived at Wind around 14:56 UT on Oct-10, with a transit time of $\sim$37 hours (assuming a CME launch time of 01:56 UT on Oct-9). Two ICMEs (or ejecta) can be identified from the magnetic field measurements following the shock. The region of rapidly increasing proton density corresponds to the sheath region behind the shock. The solar wind speed at the shock front reached $\sim$800 km s$^{-1}$ and remained relatively high, around 700 km s$^{-1}$, throughout the intervals of the leading CME. The speed profile does not decline monotonically across the interval, which indicates that the trailing ejecta maintained the speed of the leading one. Additionally, the solar wind from the coronal hole near the AR (see Figure 1f) may have played a role in maintaining the velocity. The ICME boundaries were mainly determined using the magnetic field data, in conjunction with other solar wind parameters. The two ejecta exhibit distinct magnetic configurations. The magnetic field of the leading ejecta peaked at $\sim$50 nT, which is much larger than the trailing one. For the leading ejecta, its left boundary was identified based on discontinuities in magnetic field, temperature, and density, while its right boundary was determined by discontinuities in B$_N$ and B$_R$. The magnetic field shows a sustained southward component (-B$_N$, i.e., -B$_z$ in GSE coordinates), which contributes significantly to the overall magnetic field strength. This suggests that the axis of the ejecta seems perpendicular to the ecliptic plane. However, the B$_T$ component shows multiple rotations (with its profile interrupted by a set of current-sheet-like structures around 06:00 UT on Oct-11), making the axial orientation and the ejecta chirality difficult to determine. Such a magnetic configuration indicates that the ejecta lacks clear flux-rope signatures. The proton temperature inside the leading ejecta is not depressed as usual for a CME at 1 au, likely due to the compression from the CME interactions. For the trailing ejecta, its left boundary was determined by discontinuities in B$_N$, while its right boundary was determined by discontinuities in B$_R$ and B$_T$ as well as the enhancement of solar wind speed behind the CMEs. The magnetic field strength was significantly lower than that of the leading ejecta, with almost no southward magnetic field component. Other components, however, exhibit indications of rotation, suggesting that the axis of this ICME (ejecta) is perpendicular to the ecliptic plane. Its overall structure is more consistent with a magnetic flux-rope morphology, indicating less deformation. The D$_{st}$ index exhibited the signature of a two-step geomagnetic storm. The initial decrease in the D$_{st}$ index was driven by the southward magnetic fields in the sheath region behind the shock. This was followed by a further decline to a minimum of $\sim$-333 nT, which was caused by the sustained southward field of the leading ejecta. The persistent influence of this ejecta field also led to an extended recovery phase of the D$_{st}$ profile.

\begin{figure*}[ht!]
\includegraphics[scale=0.75]{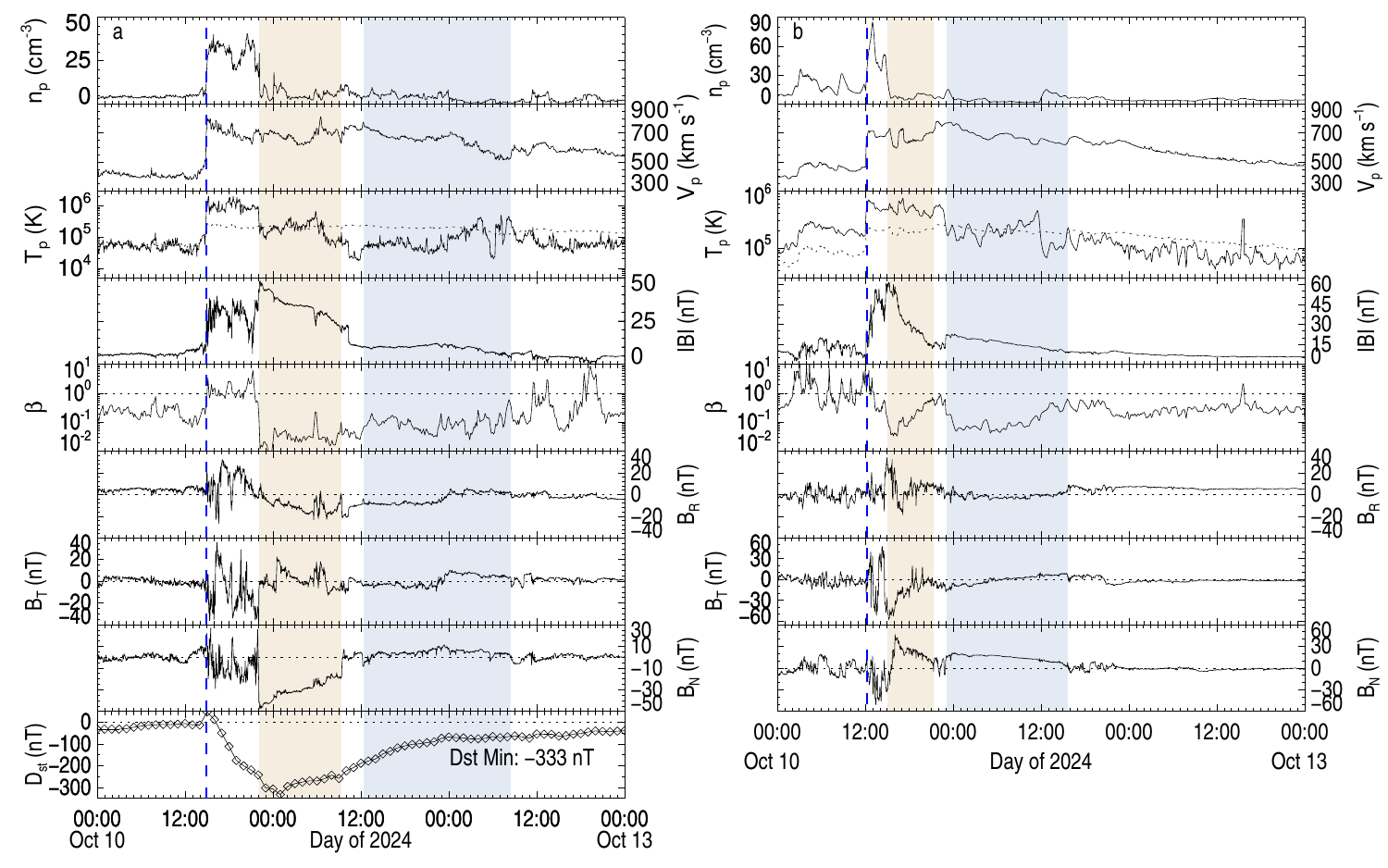}
\caption{Solar wind plasma and magnetic field parameters detected at Wind (a) and STEREO-A (b) spacecraft. (a) From top to bottom, the panels show the proton density, bulk speed, proton temperature (the dotted curve denotes the expected proton temperature from the observed speed), magnetic field strength, proton $\beta$, magnetic field components, and D$_{st}$ index. (b) Same as (a), but without the D$_{st}$ index. The yellow and blue shaded regions indicate the intervals of the two ejecta, and the blue vertical dashed lines mark the associated shocks.}
\end{figure*}

Figure 5b presents in-situ measurements from the PLASTIC instruments on STEREO-A at 0.96 au. The CME leading edge arrived $\sim$2.7 hours earlier than at Wind. The profiles of proton density, solar wind speed, and proton temperature closely resemble those observed at Wind, but the durations of these structures are noticeably shorter, which suggests that the propagation of the CMEs is more directed to STEREO-A, leading to stronger interaction. The proton temperature within the leading ejecta is higher, which is consistent with the stronger compression by stronger interaction. The magnetic field of the leading ejecta peaks at $\sim$60 nT, which is higher than that at Wind. The ICME boundaries were identified following similar criteria to those used for the Wind data. The leading CME boundary was primarily determined by the magnetic field strength profile, while for the trailing CME, the right boundary was determined by both the magnetic field and the enhancement of solar wind speed behind the CME. The intervals of the two CMEs better align with the low plasma-$\beta$ properties of CMEs. Furthermore, the trailing CME retained a magnetic field configuration similar to that observed at Wind. However, the magnetic field configuration of the leading ejecta differs significantly from that observed at Wind. In contrast, STEREO-A observes a persistently eastward B$_T$ (-B$_T$, i.e., +B$_y$). B$_N$ shows rotation but is asymmetric, with a leading field negative (-B$_z$) and trailing field positive (+B$_z$), resembling an SEN-type structure and implying left-handed chirality \citep{1998Bothmer,1998mulligan}. However, the B$_R$ component exhibits two rotations, suggesting that the structure is not an ideal flux rope. Due to the stronger magnetic field strength at STEREO-A, we expected to compare the D$_{st}$ index with that at Wind. We evaluated the D${st}$ index using two different methods. As solar wind velocity component information is currently unavailable, we use shifted plasma data from Wind as input. Due to the lack of a sustained southward CME magnetic field component, the estimated D$_{st}$ values using the formulae of \citet{2000OM} and \citet{1975Burton} only reach -141 nT and -182 nT, respectively.

\section{Conclusions and Discussions}\label{s2}

We investigated the solar and interplanetary sources of the major geomagnetic storm of 2024 October 10-11 (minimum D$_{st}$ $\sim$-333 nT)---the second strongest of Solar Cycle 25 and the fourth since 2003. Our multiviewpoint remotesensing and in-situ analysis demonstrates that this extreme geomagnetic storm was driven by a sympathetic eruption involving a quiescent filament and an active region CME, resulting in two closely interacting CMEs that manifested as a single halo structure in coronagraph images. A key insight from this event is that overlooking eruption source details could lead to underestimating the complexity of CME activity, as only one halo CME was initially reported despite the two underlying interacting structures. Similarly, in-situ measurements at multiple points were essential to confirm the successful eruption of the slow-rising quiescent filament and to reveal the enhanced southward fields from compression during the interaction.

This study aims to establish the connection between superintense geomagnetic storms and their potential solar origins. To this end, we contrast the in-situ magnetic field configurations of the two interacting CMEs (QF-CME and AR-CME) with those in their solar source regions. In-situ measurements from both spacecraft show that the trailing CME (AR-CME) exhibited ENW chirality (+B$_N$, -B$_T$ $\to$ +B$_T$ or +B$_z$, +B$_y$ $\to$ -B$_y$), which aligns well with the magnetic configuration of the inverted $\delta$-shaped warm channel in terms of both flux-rope chirality (left-handed) and axial magnetic field orientation (northward; see Figure 1). We also apply minimum and maximum variance analysis (MVA) to determine the eigenvector of the CME axis. In RTN coordinates, the derived axis is [-0.19, 0.06, 0.98] for Wind and [-0.72, 0.37, 0.58] for STEREO-A. For the trailing CME, its propagation direction relative to Earth is more head-on, so we consider its axial orientation to be consistent with the above estimate. Thus, the classification as an ENW-type CME remains plausible. The initial speeds of the CMEs were relatively high (exceeding 2000 km s$^{-1}$). Compared with the relatively weak helicity and magnetic energy injection rates, this high speed was probably related to the weak overlying magnetic confinement above the AR.

The interaction significantly enhanced the magnetic fields within the leading CME (QF-CME), which was consistently measured at both spacecraft locations. In-situ measurements from Wind and STEREO-A reveal distinct magnetic configurations with distorted flux-rope signatures, indicating deformation of the leading CME due to the interaction. The MVA yields an axis orientation of [-0.19, 0.82, 0.54] for Wind, indicating that the QF-CME axis points west by north, consistent with the orientation of the quiescent filament on the solar disk. In contrast, the axis derived from STEREO-A is [-0.39, -0.54, 0.75], showing a north-by-east orientation that differs from both the filament axial field direction and the Wind result. The more head-on perspective from STEREO-A reveals a left-handed chirality that aligns with the filament, while the overall CME axis orientation shows significant deviation. The inconsistency in CME axis orientation can be interpreted in terms of flux-rope bending during propagation. \citet{2025Al-Haddad} indicated that kink instability can drive such bending and change the flux-rope orientation. They further emphasized that CMEs frequently exhibit kinked axes, particularly in elongated and narrow flux ropes resembling the filament structure observed here. On the other hand, it should be noted that for all axial eigenvectors estimated using the MVA method, the ratio $\lambda_2$/$\lambda_3$ is around 3. This suggests that the spacecraft did not pass exactly through the center of the CMEs, or that the CMEs underwent significant deformation, resulting in a lack of clear flux-rope signatures. The latter scenario is consistent with the observed CME interaction, further reflecting the complexity of the event. Alternatively, changes in magnetic connectivity, as simulated by the CORHEL-CME model (see Appendix B for details), may offer another explanation for the observed magnetic field misalignment.

All other possible CMEs have been ruled out (see Appendix A). Therefore, we conclude that the leading CME underwent severe magnetic field compression and deformation due to the CME interaction. Consequently, a significant D$_{st}$ decrease was caused by a strong southward magnetic field at Earth, but the more head-on interaction with a stronger magnetic field strength was estimated to result in only a $\sim$100 nT D$_{st}$ decrease at STEREO-A. This highlights that while the magnetic field strength of the leading CME increases through interaction, its orientation becomes difficult to determine. Indeed, the comparison between source-region magnetic field configurations and these in-situ measurements remains tentative and requires further investigation of the coronal field configurations.

\appendix

\section{Potential Eruptions Associated with This Event}

\begin{figure*}[ht!]
\plotone{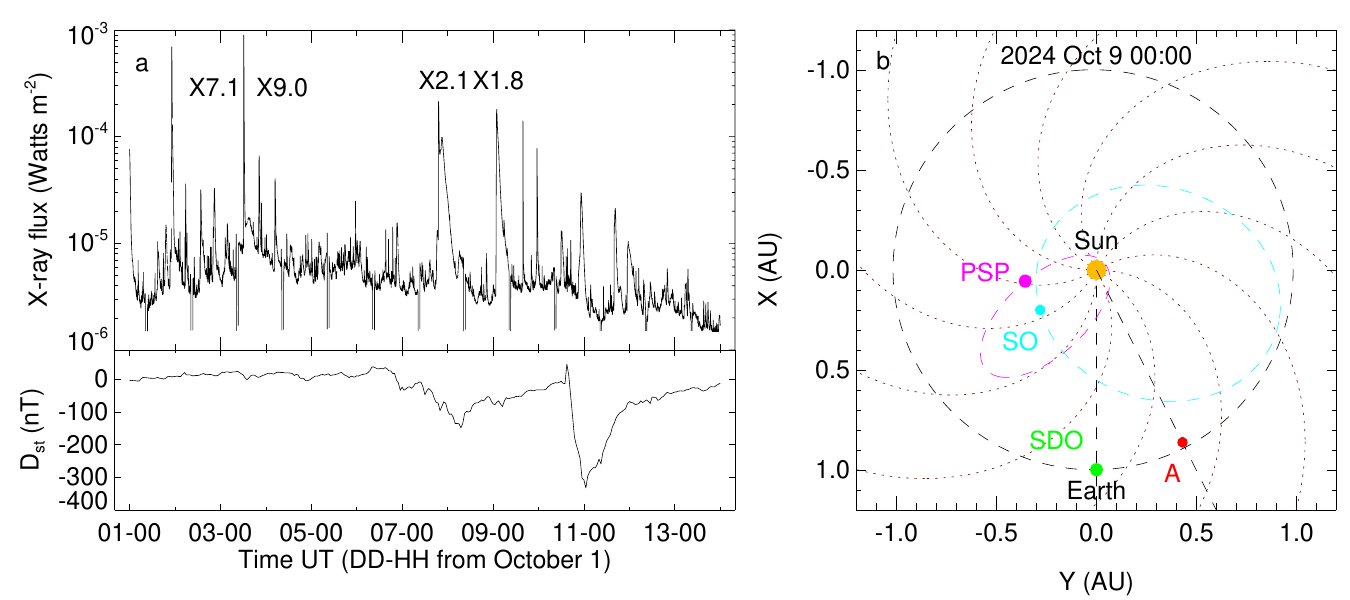}
\caption{Overview of solar eruptions in 2024 October. (a) GOES X-ray flux at 1--8 \AA~(Upper) and D$_{st}$ index (lower). (b) Positions of the spacecraft in the ecliptic plane.}
\end{figure*}

AR 13842 released extreme X7.1 and X9.0 flares accompanied by multiple CMEs (see Figure 6a), resulting in only a moderate geomagnetic storm (minimum D${st}$ $\sim$-148 nT). Despite the lower activity level, the later CME on Oct-9 led to a far more intense storm (minimum D${st}$ $\sim$ -333 nT) on Oct-11. Figure 6b shows that Earth and STEREO-A were positioned at a moderate longitudinal separation of $\sim$26.4$^\circ$. The time-elongation maps in Figure 4a reveal at least four ICMEs potentially associated with the D$_{st}$ decline on Oct-11. A comet entering the STEREO-A/HI2 FOV obscured the propagation path of these CMEs. Despite this, trends suggest possible interactions. We identified three potential eruptions associated with this geomagnetic response. The eruption closest in time to the halo CME occurred at 12:12 UT on Oct-8. GCS reconstruction indicates it is likely a backside event near the eastern limb with its source at about (E110$^\circ$, S20$^\circ$). The CME propagated toward the side and rear, making interactions with the Oct-9 CME unlikely. The CME that occurred at 06:12 UT on Oct-8 corresponds to a large filament eruption centered near (S25$^\circ$, W30$^\circ$), which erupted southward in the coronagraph. According to the WSA-ENLIL simulation results, this CME was finally directed too far south to impact Earth \footnote{\url{https://kauai.ccmc.gsfc.nasa.gov/DONKI/view/WSA-ENLIL/33848/1}}. The CME at 20:12 UT on Oct-7, occurring near the western limb at (S16$^\circ$, W85$^\circ$), was associated with AR13842. The ENLIL simulation results show that the Oct-7 CME propagated nearly westward, reducing the likelihood of interaction with the Oct-9 CME \footnote{\url{https://kauai.ccmc.gsfc.nasa.gov/DONKI/view/WSA-ENLIL/33856/1}}. Accordingly, we can rule out interference from the CMEs stated above.

\section{Magnetic field connectivity Changes by Early CME interactions}

Our research indicates that CME interactions, particularly those occurring in the lower corona, may lead to changes in magnetic field connectivity. We simulated this process using the CORHEL-CME model \citep{2018torok,2024linker}. Due to model limitations, we could not provide quantitative estimates. However, visualization of the simulation results \footnote{\url{https://ccmc.gsfc.nasa.gov/ror/results/viewrun.php?runnumber=Rui_Wang_060325_SH_5}} suggests that the QF underwent magnetic reconnection with the overlying loops, significantly altering magnetic field line connectivity. This poses challenges for accurate geomagnetic effect predictions, as our findings show that the enhanced magnetic field strength due to interactions is expected, whereas the strong southward magnetic component remains less predictable. 

\begin{acknowledgments}
We wish to thank the referee for the constructive comments that helped us improve the article. The authors thank Ting Li for her insightful comments. The research was supported by the Strategic Priority Research Program of the Chinese Academy of Sciences (No. XDB0560000), National Key R\&D Program of China (Nos. 2021YFA0718600, 2022YFF0503800), China's Space Origins Exploration Program, National Natural Science Foundation of China (NSFC, Grant No. 12073032), the fund for Key Laboratory of Space Weather of China Meteorological Administration, and the Specialized Research Fund for State Key Laboratories. H.D.H. also acknowledges support from NSFC under grant Nos. 42274201 and 42150105. X.W.Z. also acknowledges support from NSFC under grant Nos. 42204176, 42274217, and U2442202, and China Meteorological Administration ``Space Weather Monitoring and Alerting'' Key Innovation Team (CMA2024ZD01). C.C. also acknowledges support from the Research Foundation of Education Bureau of Hunan Province (No. 23B0593). S.L.M. also acknowledges support from National Key R\&D Program of China No. 2021YFA1600503 (2021YFA1600500). Z.W.Y. acknowledges support from NSFC under grant No. 42150105 and National Key R\&D Program of China under grant Nos. 2025YFF0512100 and 2021YFA0718600. C.H. acknowledges support from Scientific Research Foundation for the PhD (Huizhou University, 2023JB008). We acknowledge the use of data from SDO, STEREO, Solar Orbiter, CHASE, and ASO-S. Solar Orbiter is a space mission of international collaboration between ESA and NASA, operated by ESA. The EUI instrument was built by CSL, IAS, MPS, MSSL/UCL, PMOD/WRC, ROB, LCF/IO with funding from the Belgian Federal Science Policy Office (BELSPO/PRODEX PEA 4000134088); the Centre National d'Etudes Spatiales (CNES); the UK Space Agency (UKSA); the Bundesministerium f\"{u}r Wirtschaft und Energie (BMWi) through the Deutsches Zentrum f\"{u}r Luft- und Raumfahrt (DLR); and the Swiss Space Office (SSO). CHASE mission is supported by China National Space Administration. The ASO-S mission is supported by the Strategic Priority Research Program on Space Science, Chinese Academy of Sciences. Simulation results have been provided by the Community Coordinated Modeling Center (CCMC) at Goddard Space Flight Center through their publicly available simulation services (https://ccmc.gsfc.nasa.gov). The CORHEL-CME model was developed by Jon Linker at Predictive Science Inc.
\end{acknowledgments}

\bibliography{references}{}
\bibliographystyle{aasjournalv7}



\end{document}